\documentclass[runningheads]{llncs}
\let\llncssubparagraph\subparagraph
\let\subparagraph\paragraph

\usepackage{graphicx}
\usepackage{cite} 
\usepackage{amsmath}
\usepackage{algorithm}
\usepackage{graphicx}   
\usepackage{subcaption} 
\usepackage{booktabs} 
\usepackage{multirow} 
\usepackage{float} 
\usepackage{tikz} 
\usepackage[noend]{algpseudocode}
\usepackage{placeins} 
\usepackage{titlesec} 
\usepackage[misc]{ifsym} 

\let\subparagraph\llncssubparagraph

\newcommand{\SHIFT}{\text{SHIFT}}
\newcommand{\HASH}{\text{HASH}}
\newcommand{\POS}{\text{POS}}
\newcommand{\fp}{\textit{\text{fp}}}
\newcommand{\idx}{\textit{\text{index}}}
\newcommand{\red}{\textcolor{black}}

\newcommand\blfootnote[1]{%
  \begingroup
  \renewcommand\thefootnote{}\footnote{#1}%
  \addtocounter{footnote}{-1}%
  \endgroup
}

\titlespacing{\section}{0pt}{9pt}{5pt}
\titlespacing{\subsection}{0pt}{9pt}{4pt}
\begin{document}
\title{Fast Multiple Pattern Cartesian Tree Matching}
%
%
\author{Geonmo Gu\inst{1}, Siwoo Song\inst{1}, Simone Faro\inst{2}, Thierry Lecroq\inst{3}, \\and Kunsoo Park\inst{1}\textsuperscript{(\Letter)}}
\authorrunning{G. Gu et al.}
%
\institute{Seoul National University, Seoul, Korea\\ \email{\{gmgu,swsong,kpark\}@theory.snu.ac.kr} \and
University of Catania, Catania, Italy\\ \email{faro@dmi.unict.it} \and Normandie University, Rouen, France\\ \email{thierry.lecroq@univ-rouen.fr}}
\maketitle              
\begin{abstract}
Cartesian tree matching is the problem of finding all substrings in a given text which have the same Cartesian trees as that of a given pattern. In this paper, we deal with Cartesian tree matching for the case of multiple patterns. We present two fingerprinting methods, i.e., the parent-distance encoding and the binary encoding. By combining an efficient fingerprinting method and a conventional multiple string matching algorithm, we can efficiently solve multiple pattern Cartesian tree matching. We propose three practical algorithms for multiple pattern Cartesian tree matching based on the Wu-Manber algorithm, the Rabin-Karp algorithm, and the Alpha Skip Search algorithm, respectively. In the experiments we compare our solutions against the previous algorithm \cite{park2019cartesian}. Our solutions run faster than the previous algorithm as the pattern lengths increase. Especially, our algorithm based on Wu-Manber runs up to 33 times faster.

\end{abstract}
\section{Introduction}
    
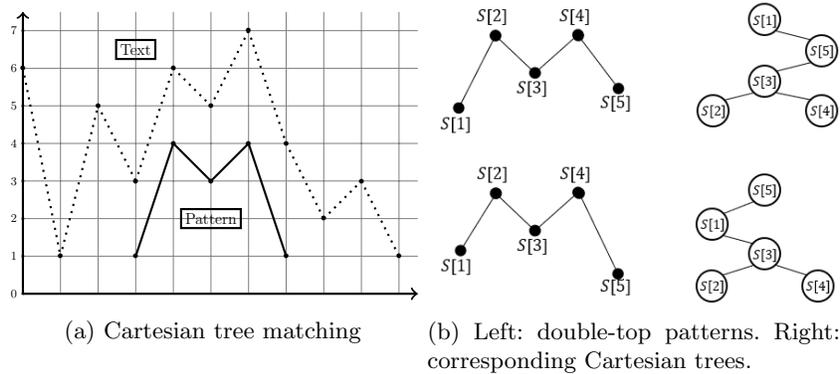
\begin{figure}[t]
    \centering
    \begin{subfigure}[b]{0.45\textwidth}
        \centering
        \begin{tikzpicture}[thick,scale=0.5, every node/.style={transform shape}]
            \draw[step=1cm,gray,very thin] (0,0) grid (10.5,7.5); 
            \draw[thick,->] (0,0) -- (10.5,0); 
            \draw[thick,->] (0,0) -- (0,7.5); 
            \foreach \y in {0,1,2,3,4,5,6,7} 
                \draw (1pt,\y cm) -- (-1pt,\y cm) node[anchor=east] {$\y$};
            \draw[dotted] (0,6) -- (1,1);
            \draw[dotted] (1,1) -- (2,5);
            \draw[dotted] (2,5) -- (3,3);
            \draw[dotted] (3,3) -- (4,6);
            \draw[dotted] (4,6) -- (5,5);
            \draw[dotted] (5,5) -- (6,7);
            \draw[dotted] (6,7) -- (7,4);
            \draw[dotted] (7,4) -- (8,2);
            \draw[dotted] (8,2) -- (9,3);
            \draw[dotted] (9,3) -- (10,1);
            \draw (3,1) -- (4,4);
            \draw (4,4) -- (5,3);
            \draw (5,3) -- (6,4);
            \draw (6,4) -- (7,1);
            
            \foreach \Point in {(0,6), (1,1), (2,5), (3,3), (4,6), (5,5), (6,7), (7,4), (8,2), (9,3), (10,1), (3,1), (4,4), (5,3), (6,4), (7,1)}
                \node at \Point {\textbullet};
            
            \node[draw] at (3,6.5) {\large Text};
            \node[draw] at (5,2) {\large Pattern};
            
        \end{tikzpicture}
        \caption{Cartesian tree matching}
        \vspace{4mm}
        \label{fig:example_text_pattern}
    \end{subfigure}
    \begin{subfigure}[b]{0.45\textwidth}
        \centering
        \includegraphics[width=\linewidth]{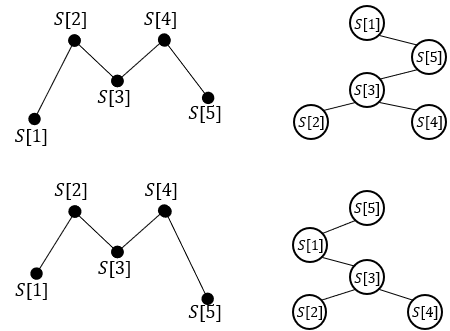}
        \caption{Left: double-top patterns. Right: corresponding Cartesian trees.}
        \label{fig:cartesian_trees_for_doubletop}
    \end{subfigure}
    \vspace{-3mm}
    \caption{Cartesian tree matching: multiple Cartesian trees are required for the double-top pattern.}
    \label{fig:motivation}
\end{figure}

    \blfootnote{\red{Gu, Song, and Park were supported by Collaborative Genome Program for Fostering New Post-Genome industry  through the National Research Foundation of Korea(NRF) funded by the Ministry of  Science ICT and Future Planning (No. NRF-2014M3C9A3063541).}}
    Cartesian tree matching is the problem of finding all substrings in a given text which have the same Cartesian trees as that of a given pattern. For instance, given text $T=(6, 1, 5, 3, 6, 5, 7, 4, 2, 3, 1)$ and pattern $P=(1, 4, 3, 4, 1)$ in Figure \ref{fig:example_text_pattern}, $P$ has the same Cartesian tree as the substring $(3, 6, 5, 7, 4)$ of $T$. Among many generalized matchings, Cartesian tree matching is analogous to order-preserving matching\red{\cite{kim2014order, kubica2013linear, chhabra2014order, ganguly2016space}} in the sense that they deal with relative order between numbers. Accordingly, both of them can be applied to time series data such as stock price analysis, but Cartesian tree matching can be sometimes more appropriate than order-preserving matching in finding patterns \cite{park2019cartesian}.
        
    In this paper, we deal with Cartesian tree matching for the case of multiple patterns. Although finding multiple different patterns is interesting by itself, multiple pattern Cartesian tree matching can be applied in finding one meaningful pattern when the meaningful pattern is represented by multiple Cartesian trees: Suppose we are looking for the double-top pattern \cite{liu2007automatic}. Two Cartesian trees in Figure \ref{fig:cartesian_trees_for_doubletop} are required to identify the pattern, where the relative order between $S[1]$ and $S[5]$ causes the difference. In general, the more complex the pattern is, the more Cartesian trees having the same lengths are required. (e.g., the head-and-shoulder pattern \cite{liu2007automatic} requires four Cartesian trees.)
    
   Recently, Park et al.~\cite{park2019cartesian} introduced (single pattern) Cartesian tree matching, multiple pattern Cartesian tree matching, and Cartesian tree indexing with their respective algorithms. They proposed the parent-distance representation that has a one-to-one mapping with Cartesian trees, and gave linear-time solutions for the problems, utilizing the representation and existing string algorithms, i.e., KMP algorithm, Aho-Corasick algorithm, and suffix tree construction algorithm. Song et al.~\cite{song2019fast} proposed new representations about Cartesian trees, and proposed practically fast algorithms for Cartesian tree matching based on the framework of filtering and verification. 

    Extensive works have been done to develop algorithms for multiple pattern matching, which is one of the fundamental problems in computer science \cite{song2008memory, hua2009variable, liao2013intrusion}.
    Aho and Corasick \cite{aho1975efficient} presented a linear-time algorithm based on an automaton. Commentz-Walter \cite{commentz1979string} presented an algorithm that combines the Aho-Corasick algorithm and the Boyer-Moore technique \cite{boyer1977fast}. Crochemore et al.~\cite{crochemore1999fast} proposed an algorithm that combines the Aho-Corasick automaton and a Directed Acyclic Word Graph, which runs linear in the worst case and runs in  $O((n/m)\log m)$ time in the average case, where $m$ is the length of the shortest pattern. Rabin and Karp \cite{karp1987efficient} proposed an algorithm that runs linear on average and $O(nM)$ in the worst case, where $M$ is the sum of lengths of all patterns. Charras et al.~\cite{charras1998very} proposed an algorithm called Alpha Skip Search, which can efficiently handle both single pattern and multiple patterns. Wu and Manber \cite{wu1994fast} presented an algorithm that uses an extension of the Boyer-Moore-Horspool technique. 
    
        
    In this paper we present practically fast algorithms for multiple pattern Cartesian tree matching. 
    We present three algorithms based on Wu-Manber, Rabin-Karp, and Alpha Skip Search. All of them use the filtering and verification approach, where filtering relies on efficient fingerprinting methods of a string. Two fingerprinting methods are presented, i.e., the parent-distance encoding and the binary encoding. By combining an efficient fingerprinting method and a conventional multiple string matching algorithm, we can efficiently solve multiple pattern Cartesian tree matching. In the experiments we compare our solutions against the previous algorithm \cite{park2019cartesian} which is based on the Aho-Corasick algorithm. Our solutions run faster than the previous algorithm. Especially, our algorithm based on Wu-Manber runs up to 33 times faster. 
    
\section{Problem Definition}
    \subsection{\red{Notation}}

        A \textit{string} is a sequence of characters drawn from an alphabet $\Sigma$, which is a set of integers. We assume that a comparison between any two characters can be done in constant time. For a string $S$, $S[i]$ represents the $i$-th character of $S$, and $S[i..j]$ represents the substring of $S$ starting from $i$ and ending at $j$.
        
        A \textit{Cartesian tree} \cite{vuillemin1980unifying} is a binary tree derived from a string. Specifically, the Cartesian tree $CT(S)$ for a string $S$ can be uniquely defined as follows:
        \begin{itemize}
            \item If $S$ is an empty string, $CT(S)$ is an empty tree.
            \item If $S$ is not empty and $S[i]$ is the minimum value in $S[1..n]$, $CT(S)$ is the tree with $S[i]$ as the root, $CT(S[1..i-1])$ as the left subtree, and $CT(S[i+1..n])$ as the right subtree. If there is more than one minimum value, we choose the leftmost one as the root.
        \end{itemize}
        
        Given two strings $T[1..n]$ and $P[1..m]$, where $m \leq n$, we say that $P$ \textit{matches} $T$ at position $i$ if $CT(T[i-m+1..i])=CT(P[1..m])$. For example, given \red{$T=(6, 1, 5, 3, 6, 5, 7, 4, 2, 3, 1)$ and $P=(1,4,3,4,1)$ in Figure \ref{fig:example_text_pattern}}, $P$ matches $T$ at position \red{8}. We also say that $T[\red{4..8}]$ is \textit{a match} of $P$ in $T$.
        

        \textit{Cartesian tree matching} is the problem of finding all the matches in the text which have the same Cartesian trees as a given pattern.
        \begin{definition}
            (Cartesian tree matching \cite{park2019cartesian}) Given two strings text $T[1..n]$ and pattern $P[1..m]$, find every $m \leq i \leq n$ such that $CT(T[i-m+1..i])=CT(P[1..m])$.
        \end{definition}
        
    \subsection{Multiple Pattern Cartesian Tree Matching}

        Cartesian tree matching can be extended to the case of multiple patterns. \textit{Multiple pattern Cartesian tree matching} is the problem of finding all the matches in the text which have the same Cartesian trees as at least one of the given patterns. 
        \begin{definition}
            (Multiple pattern Cartesian tree matching \cite{park2019cartesian}) Given a text $T[1..n]$ and patterns $P_1[1..m_1],P_2[1..m_2], ..., P_k[1..m_k]$, find every position in the text which matches at least one pattern, i.e., it has the same Cartesian tree as that of at least one pattern.
        \end{definition}

\section{Fingerprinting Methods}
\label{sec:fingerprinting}

    Fingerprinting is a technique that maps a string to a much shorter form of data, such as a bit string or an integer. 
    In Cartesian tree matching, we can use fingerprints to filter out unpromising matching positions with low computational cost.
    
    In this section we introduce two fingerprinting methods, i.e., the parent-distance encoding and the binary encoding, for the purpose of representing information about Cartesian tree as an integer. The two encodings make use of the parent-distance representation and the binary representation, respectively, both of which are strings that represent Cartesian trees.

    \subsection{Parent-distance Encoding}

        In order to represent Cartesian trees efficiently, Park et al.~proposed the \textit{parent-distance representation} \cite{park2019cartesian}\red{, which is another form of the all nearest smaller values \cite{berkman1993optimal}.}
        \begin{definition}
            (Parent-distance representation) Given a string S[1..n], the parent-distance representation of S is an integer string PD(S)[1..n], which is defined as follows:
            \begin{equation}
                PD(S)[i] = 
                \begin{cases}
                    i-\max_{1 \leq j < i}\{j : S[j] \leq S[i]\} & \mbox{if such j exists}\\
                    0 & \mbox{otherwise}
                \end{cases}
            \end{equation}
        \end{definition}
        
        Intuitively, $PD(S)[i]$ stores the distance between $S[i]$ and the parent of $S[i]$ in $CT(S[1..i])$. For example, the parent-distance representation of string $S=(11,14,13,15,12)$ is $PD(S)=(0,1,2,1,4)$, where $PD(S)[3]=3-1=2$ stores the distance between $S[3]$ and $S[1]$ ($S[1]$ is the parent of $S[3]$ in $CT(S[1..3])$). The parent-distance representation has a one-to-one mapping to the Cartesian tree \cite{park2019cartesian}, and so if two strings have the same parent-distance representations, the two strings also have the same Cartesian trees. The parent-distance representation of a string can be computed in linear time \cite{park2019cartesian}. Note that $PD(S)[i]$ holds a value between $0$ to $i-1$ by definition, and $PD(S)[1]=0$ at all times.
        
        With the parent-distance representation, we can define a fingerprint encoding function that maps a string to an integer, using the factorial number system \cite{knuth2014art}.
        \begin{definition}
            (Parent-distance Encoding) Given a string $S[1..n]$, \red{the} encoding function $f(S)$, which maps $S$ into an integer within the range $[0..n!-1]$, is defined as follows:
            \begin{equation}
                f(S)=\sum_{i=2}^n(PD(S)[i])\cdot(i-1)!.
            \end{equation}
        \end{definition}
        The parent-distance encoding maps a string into \red{a} unique integer according to its parent-distance representation. That is, given two strings $S_1$ and $S_2$, $CT(S_1)=CT(S_2)$ if and only if $f(S_1)=f(S_2)$. This is because if $PD(S_1) \neq PD(S_2)$ then $f(S_1) \neq f(S_2)$ due to the fact that $PD(S)[i]<i$. The encoding function $f(S[1..n])$ can be computed in $O(n)$ time, since $PD(S)$ can be computed in linear time. For a long string, the fingerprint may not fit in a word size, so we select a prime number by which we divide the fingerprint, and use the residue instead of the actual fingerprint. A similar encoding function was used to solve the multiple pattern order-preserving matching problem \cite{han2015fast}.
    
    \subsection{Binary Encoding}

        For order-preserving matching, the representation of a string as a binary string is first presented by Chhabra and Tarhio \cite{chhabra2014order}. Recently, Song et al.~make use of the \textit{binary representation} for Cartesian tree matching as follows \cite{song2019fast}.
        \begin{definition}
            (Binary representation) Given an $n$-length string $S$, binary representation $\beta(S)$ of length $n-1$ is defined as follows: for $1 \leq i \leq n-1$,
            \begin{equation}
                \beta(S)[i] = 
                \begin{cases}
                    1 & \mbox{if } S[i] \leq S[i+1] \\
                    0 & \mbox{otherwise.}
                \end{cases}
            \end{equation}
        \end{definition}
        
        Given two strings $S_1[1..n]$ and $S_2[1..n]$, the binary representations $\beta(S_1)$ and $\beta(S_2)$ are the same if the Cartesian trees $CT(S_1)$ and $CT(S_2)$ are the same \cite{song2019fast}. Obviously, the Cartesian tree has a many-to-one mapping to the binary representation. Thus, two strings whose binary representations are the same may not have the same Cartesian trees, but two strings whose Cartesian trees are the same have the same binary representations.

        A fingerprint encoding function $f(S)$ can be defined using the binary representation. 
        \begin{definition}
            (Binary Encoding) Given a string $S[1..n]$, encoding function $f(S)$, which maps $S$ into an integer within the range $[0..2^{n-1}-1]$, is defined as follows:
            \begin{equation}
                f(S)=\sum_{i=1}^{n-1}(\beta(S)[i]\cdot2^{n-1-i}).
            \end{equation}
        \end{definition}
        Since $f(S)$ is a polynomial, it can be efficiently computed in linear time using Horner's rule \cite{cormen2001introduction}. Moreover, a fingerprint computed by the binary encoding can be reused when two strings overlap, which will be discussed in Appendix \ref{subsec:reusefingerprint}. Like the parent-distance encoding, in case the fingerprint does not fit in a word size, we select a prime number by which we divide the fingerprint, and use the residue instead of the actual fingerprint.
        
\section{Fast Multiple Pattern Cartesian Tree Matching Algorithms}
\label{sec:fastalgorithms}

    In this section we introduce three algorithms for multiple pattern Cartesian tree matching. Each of them consists of preprocessing and search. In the preprocessing step, hash tables are built using fingerprints of patterns. In the search step, the filtering and verification approach is adopted. To filter out unpromising matching positions, a fingerprinting method is applied to either length-$m$ substrings of the text, where $m$ is the length of the shortest pattern, or much shorter length-$b$ substrings of the text (we will discuss how to set $b$ in Section \ref{subsec:blocksize}). Then each candidate pattern is verified by an efficient comparison method (see Appendix \ref{subsec:verification}). 
    
    \subsection{Algorithm Based on Wu-Manber}

        \begin{algorithm}[t]
            \caption{Algorithm based on Wu-Manber}
            \label{alg:Wu-Manber}
            \begin{algorithmic}[1]
                \State $\textbf{input: }\text{text } T[1..n]\text{ and patterns } P_1[1..m_1], P_2[1..m_2], ..., P_k[1..m_k]$
                \State $\textbf{output: }\text{every position in } T \text{ that matches at least one of the patterns}$
                
                \Procedure{Preprocessing}{}
                    \State $m \gets \text{min}(m_1,m_2,...,m_k)$
                    \State $b \gets {\log_2 (km)}$
                    \State Initialize each entry of $\SHIFT$ to $m-b+1$
                    \For{$i \gets 1$ to $k$}
                        \For{$j \gets b$ to $m-1$}
                            \State {$\fp$ $\gets$ $f(P_i[j-b+1..j])$}
                            \If{$\SHIFT[\fp] > m-j$}
                                \State $\SHIFT[\fp] \gets m-j$
                            \EndIf
                        \EndFor
                        
                        \State $\fp$ $\gets$ $f(P_i[m-b+1..m])$
                        \State $\HASH[\fp].add(i)$
                    \EndFor
                \EndProcedure
                    
                \Procedure{Search}{}
                    \State $\idx \gets m$
                    \While{$\idx \leq n$}
                        \State $\fp \gets f(T[\idx -b+1..\idx])$
                        \For{$i \in \HASH[\fp]$}
                            \If{$P_i$ matches $T[\idx -m+1..\idx -m+m_i]$}
                                \State output $\idx -m+m_i$
                            \EndIf
                        \EndFor
                        \State $\idx \gets \idx + \SHIFT[\fp]$
                    \EndWhile
                \EndProcedure
            \end{algorithmic}
        \end{algorithm}
    
        Algorithm \ref{alg:Wu-Manber} shows the pseudo-code of an algorithm for multiple pattern Cartesian tree matching based on the Wu-Manber algorithm \cite{wu1994fast}. The algorithm uses two hash tables, HASH and SHIFT. Both tables use a fingerprint of length-$b$ string, called a \emph{block}. Either the parent-distance encoding or the binary encoding is used to compute the fingerprint. Given patterns $P_1,P_2,...,P_k$, let $m$ be the length of the shortest pattern. $\HASH$ maps a fingerprint $\fp$ of a block to the list of patterns $P_i$ such that the fingerprint of the last block in $P_i$'s length-$m$ prefix is the same as $\fp$. For a block $B[1..b]$ and a fingerprint encoding function $f$, HASH is defined as follows:
        \begin{equation}
            \HASH[f(B)]=\{i : f(P_i[m-b+1..m]) = f(B), 1\leq i \leq k\}
        \end{equation} 
        $\SHIFT$ maps a fingerprint $\fp$ of a block to the amount of a valid shift when the block appears in the text. The shift value is determined by the rightmost occurrence of a block in terms of the fingerprint among length-$(m-1)$ prefixes of the patterns. For a block $B[1..b]$ and a fingerprint encoding function $f$, we define the rightmost occurrence $r_B$ as follows:
        \begin{equation}
            r_B = 
            \begin{cases}
                \max_{b \leq j \leq m-1}\{j : f(P_i[j-b+1..j]) = f(B), 1 \leq i \leq k\} & \mbox{if such $j$ exists}\\
                0 & \mbox{otherwise}
            \end{cases}
        \end{equation}
        Then SHIFT is defined as follows:
        \begin{equation}
            \SHIFT[f(B)] = m - r_B
        \end{equation}
        
        In the preprocessing step, we build $\HASH$ and $\SHIFT$ (as described in Algorithm \ref{alg:Wu-Manber}). In the search step, we scan the text from left to right, computing the fingerprint of a length-$b$ substring of the text to get a list of patterns from $\HASH$. Let $\idx$ be the current scanning position of the text. We compute fingerprint $\fp$ of $T[\idx-b+1..\idx]$, and get a list of patterns in the entry $\HASH[\fp]$. If the list is not empty, each pattern is verified by an efficient comparison method (see Appendix \ref{subsec:verification}). Consider $P_i[1..m_i]$ in the list. The comparison method verifies whether $P_i$ matches $T[\idx-m+1..\idx-m+m_i]$. After verifying all patterns in the list, the text is shifted by $\SHIFT[\fp]$.

        The worst case time complexity of Algorithm \ref{alg:Wu-Manber} is $O((M+b)n)$, where $M$ is the total pattern length, $b$ is the block size, and $n$ is the length of the text (consider $T=1^n$ and the patterns of which prefixes are $1^m$). On the other hand, the best case time complexity of Algorithm \ref{alg:Wu-Manber} is $O({bn \over{m-b}})$.

    \subsection{Algorithm Based on Rabin-Karp}
        
        Algorithm \ref{alg:Rabin-Karp} in Appendix shows the pseudo-code of an algorithm for multiple pattern Cartesian tree matching based on the Rabin-Karp algorithm \cite{karp1987efficient}. The algorithm uses one hash table, namely $\HASH$. HASH is similarly defined as in Algorithm \ref{alg:Wu-Manber} except that we consider length-$m$ prefixes instead of blocks and we use only binary encoding for fingerprinting. For a string $S[1..m]$ and the binary encoding function $f$, HASH is defined as follows:
        \begin{equation}
            \HASH[f(S)]=\{i : f(P_i[1..m]) = f(S), 1\leq i \leq k\}
        \end{equation}
        
        In the preprocessing step, we build $\HASH$. In the search step, we shift one by one, and compute the fingerprint of a length-$m$ substring of the text to get candidate patterns by using HASH. Again, each candidate pattern is verified by an efficient comparison method.
        
        Given a fingerprint at position $i$ of the text, the next fingerprint at position $i+1$ can be computed in constant time if we use the binary encoding as a fingerprinting method. Let the former fingerprint be $\fp_i=f(T[i-m+1..i])$ and the latter one be $\fp_{i+1}=f(T[i-m+2..i+1])$. Then,
        \begin{equation}
        \label{eq:successivefinger}
            \fp_{i+1}=2(\fp_{i}-2^{m-2}\beta(T)[i-m+1]) + \beta(T)[i]
        \end{equation}
        Subtracting $2^{m-2}\beta(T)[i-m+1]$ removes the leftmost bit from $\fp_i$, multiplying the result by 2 shifts the number to the left by one position, and adding $\beta(T)[i]$ brings in the appropriate rightmost bit. 
        
        The worst case time complexity of Algorithm \ref{alg:Rabin-Karp} is $O(Mn)$ (consider $T=1^n$ and patterns of which prefixes are $1^m$). The best case time complexity is $O(n)$ since fingerprint $f_i$ at position $i$, $m+1 \leq i \leq n$, can be computed in $O(1)$ time using Equation (\ref{eq:successivefinger}).
        
    \subsection{Algorithm Based on Alpha Skip Search}
    
        Algorithm \ref{alg:AlphSkipSearch} in Appendix shows the pseudo-code of an algorithm for multiple pattern Cartesian tree matching based on Alpha Skip Search \cite{charras1998very}. Recall that a length-$b$ string is called a block. The algorithm uses a hash table POS that maps the fingerprint of a block to a list of occurrences in all length-$m$ prefixes of the patterns. Either the parent-distance encoding or the binary encoding is used for fingerprinting. For a block $B[1..b]$ and a fingerprint encoding function $f$, POS is defined as follows:
        \begin{equation}
            \POS[f(B)] = \{(i,j) : f(P_i[j-b+1..j])=f(B), 1\leq i\leq k, b \leq j \leq m \}
        \end{equation}
        
        In the preprocessing step, we build $\POS$. In the search step, we scan the text from left to right, computing the fingerprint of a length-$b$ substring of the text to get the list of pairs $(i,j)$, meaning that the fingerprint of $P_i[j-b+1..j]$ is the same as that of the substring of the text. Verification using an efficient comparison method is performed for each pair in the list. Note that the algorithm always shifts by $m-b+1$.
        
        The worst case time complexity of Algorithm \ref{alg:AlphSkipSearch} is $O((M+b)n)$, where $M$ is the total pattern length, $b$ is the block size, and $n$ is the length of the text (consider $T=1^n$ and patterns of which prefixes are $1^m$). On the other hand, the best case time complexity of Algorithm \ref{alg:AlphSkipSearch} is $O({bn \over{m-b}})$ since the algorithm always shifts by $m-b+1$.

    \subsection{Selecting the Block Size}
    \label{subsec:blocksize}

        The size of the block affects the running time of Algorithms \ref{alg:Wu-Manber} and \ref{alg:AlphSkipSearch}. A longer block size leads to a lower probability of candidate pattern occurrences, so it decreases verification time. On the other hand, a longer block size increases the overhead required for computing fingerprints. Thus, it is important to set a block size appropriate for each algorithm.
        
        In order to set a block size, we first study the matching probability of two strings, in terms of Cartesian trees. Assume that numbers are independent and identically distributed, and there are no identical numbers within any length-$n$ string. 
        \begin{lemma}
        \label{lem:matchingprobability}
            Given two strings $S_1[1..n]$ and $S_2[1..n]$, the probability $p(n)$ that $S_1$ and $S_2$ have the same Cartesian tree can be defined by the recurrence formula, where $p(0)=1$ and $p(1)=1$, as follows:
            \begin{equation}
                p(n) = {p(0)p(n-1)+p(1)p(n-2)+ \dots +p(n-1)p(0) \over{n^2}}
            \end{equation}
        \end{lemma}
        
        We have the following upper bound on the matching probability.
        \begin{theorem}
        \label{thm:matchingprobability}
            Assume that numbers are independent and identically distributed, and there are no identical numbers within any length-$n$ string. Given two strings $S_1[1..n]$ and $S_2[1..n]$, the probability that the two strings match, in terms of Cartesian trees, is at most ${1\over2^{n-1}}$, i.e., $p(n) \leq {1\over2^{n-1}}$. 
        \end{theorem}
        
        We set the block size $b={\log_2 (km)}$ if $\log_2(km) \leq m$; otherwise we set $b=m$, where $k$ is the number of patterns and $m$ is the length of the shortest pattern, in order to get a low probability of match and a relatively short block size with respect to $m$. By Theorem \ref{thm:matchingprobability}, if we set $b=\log_2(km)$, $p(b) \leq {2\over{km}}$. 

\section{Experiments}
    
    We conduct experiments to evaluate the performances of the proposed algorithms against the previous algorithm. We compare algorithms based on Aho-Corasick (AC) \cite{park2019cartesian}, Wu-Manber (WM), Rabin-Karp (RM), and Alpha Skip Search (AS). By default, all our algorithms use optimization techniques introduced in Appendix \ref{sec:optimization}, except the min-index filtering method which is evaluated in the experiments. Particularly, in order to compare the fingerprinting methods and see the effect of min-index filtering method, we compare variants of our algorithms. The following algorithms are evaluated.
    \begin{itemize}
        \item AC: multiple Cartesian tree matching algorithm based on Aho-Corasick \cite{park2019cartesian}.
        \item WMP: algorithm based on Wu-Manber that uses the parent-distance encoding as a fingerprinting method.
        \item WMB: algorithm based on Wu-Manber that uses the binary encoding as a fingerprinting method. The algorithm reuses fingerprints when adjacent blocks overlap $b-1$ characters (i.e., when the text shifts by one position), where $b$ is the block size.
        \item WMBM: WMB that exploits additional min-index filtering in Appendix \ref{subsec:minfiltering}.
        \item RK: algorithm based on Rabin-Karp that uses the binary encoding as a fingerprinting method.
        \item ASB: algorithm based on Alpha Skip Search that uses the binary encoding as a fingerprinting method. The algorithm reuses fingerprints when adjacent blocks overlap $b-1$ characters.
    \end{itemize}
    All algorithms are implemented in C++. Experiments are conducted on a machine with Intel Xeon E5-2630 v4 2.20GHz CPU and 128GB memory running CentOS Linux.
    
    The total time includes the preprocessing time for building data structures and the search time. To evaluate an algorithm, we run it 100 times and measure the average total time in milliseconds.
    
    We randomly build a text of length 10,000,000 where the alphabet size is 1,000. A pattern is extracted from the text at a random position.

    \subsection{Evaluation on the Equal Length Patterns}
    \label{subsec:exp_rnd_equal}
    
        \begin{figure}[h!]
    \centering

    \begin{subfigure}[b]{0.9\textwidth}
        \includegraphics[width=\linewidth]{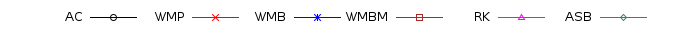}
        \label{fig:exp_pattern_length_label}
    \end{subfigure}

    \begin{subfigure}[b]{0.45\textwidth}
        \includegraphics[width=\linewidth]{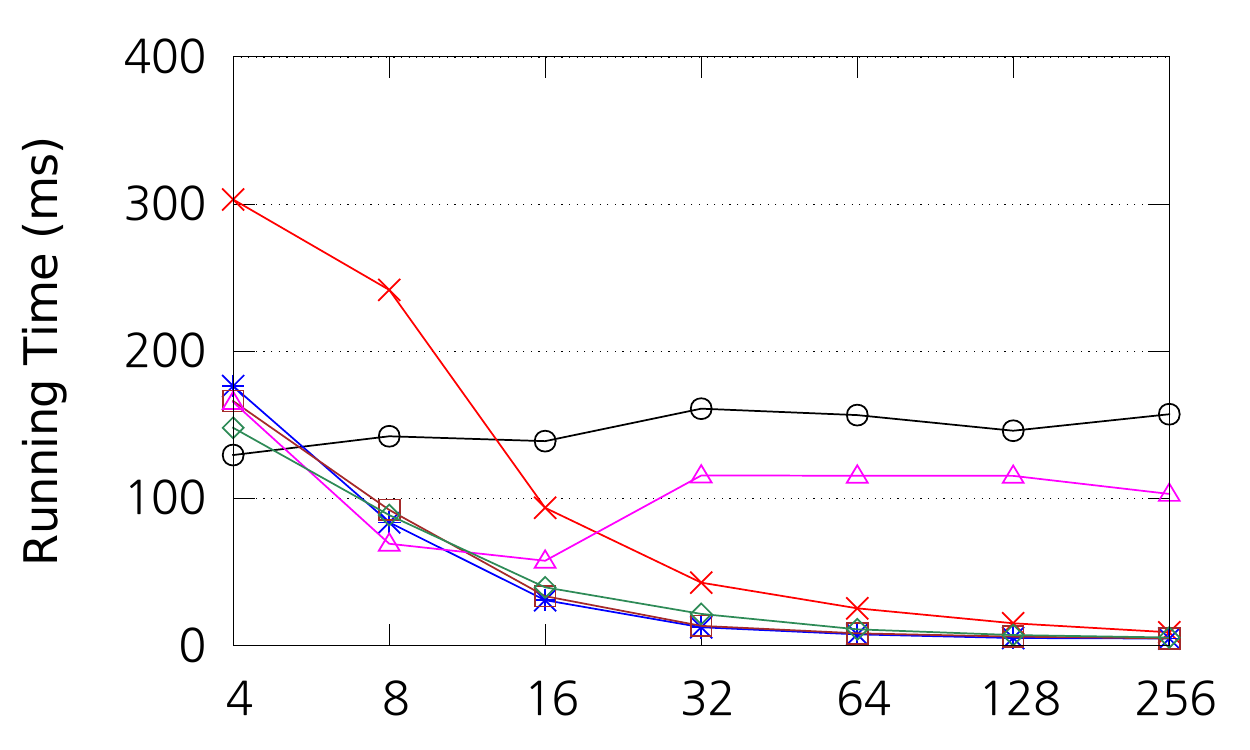}
        \vspace{-5mm}
        \caption{$k=10$}
        \label{fig:exp_equal_length_k10}
    \end{subfigure}
    \begin{subfigure}[b]{0.45\textwidth}
        \includegraphics[width=\linewidth]{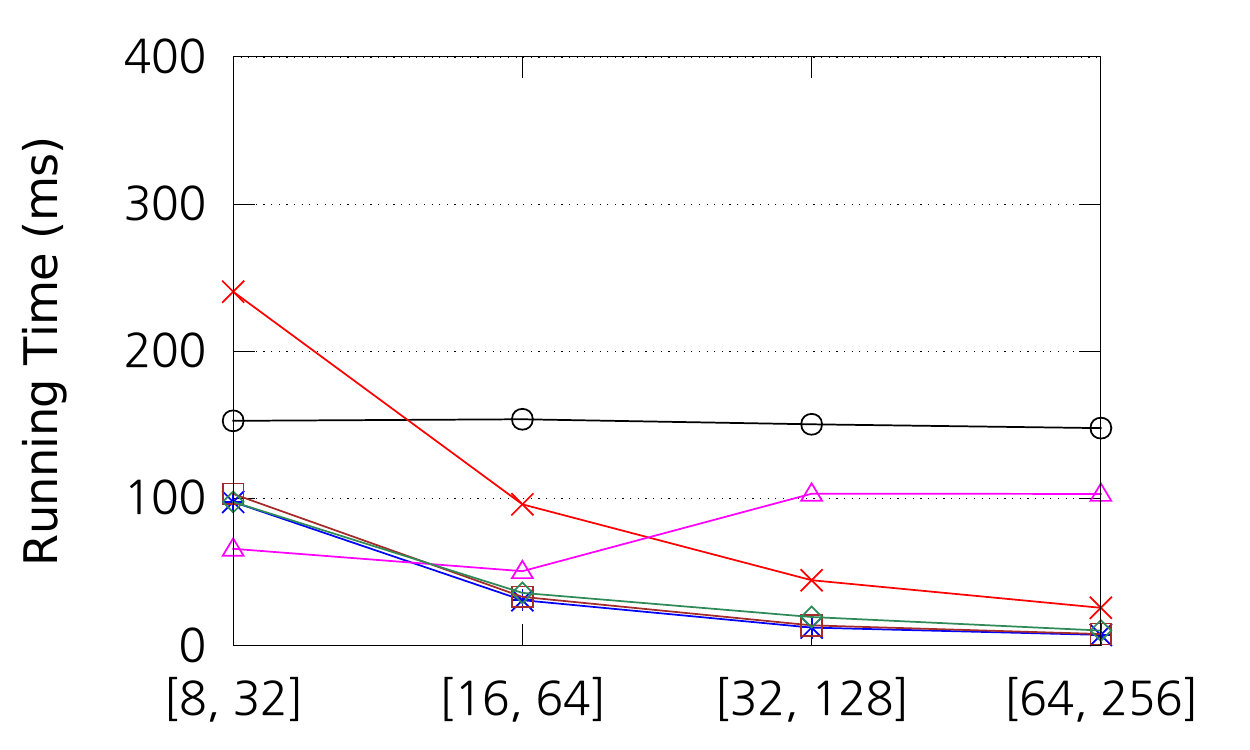}
        \vspace{-5mm}
        \caption{$k=10$}
        \label{fig:exp_diff_length_k10}
    \end{subfigure}
    
    \begin{subfigure}[b]{0.45\textwidth}
        \includegraphics[width=\linewidth]{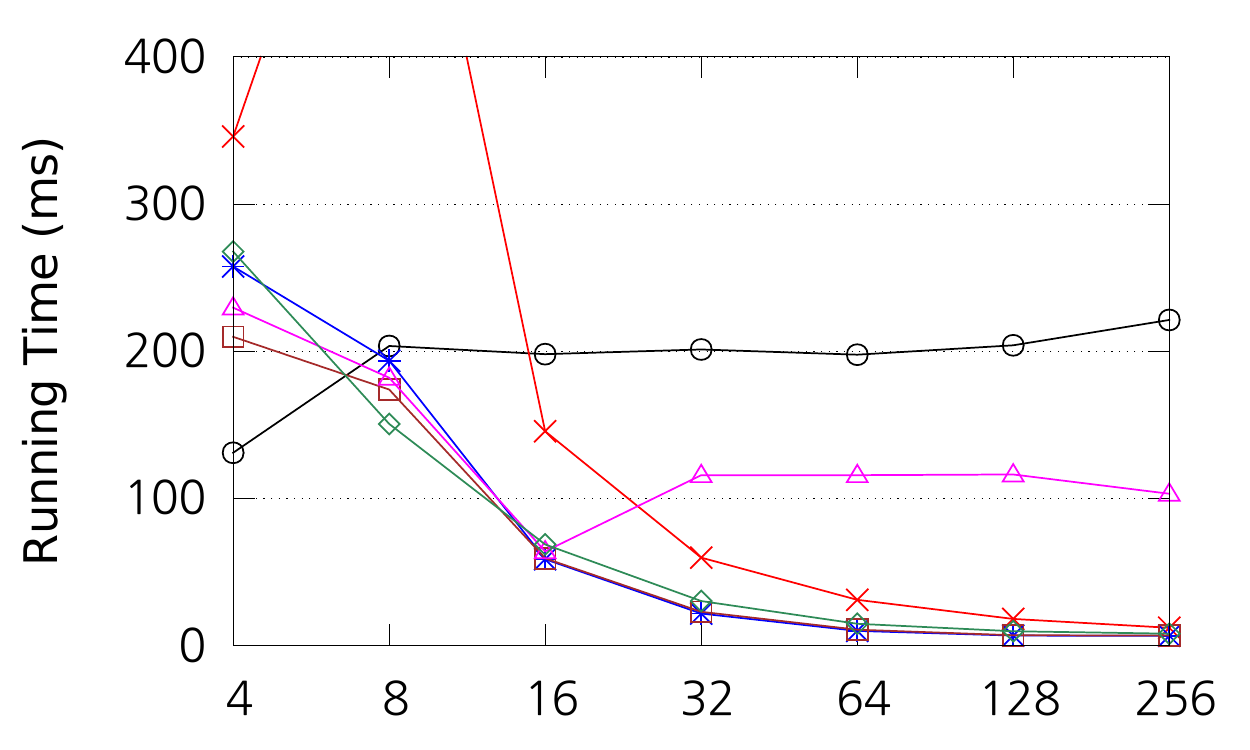}
        \vspace{-5mm}
        \caption{$k=50$}
        \label{fig:exp_equal_length_k50}
    \end{subfigure}
    \begin{subfigure}[b]{0.45\textwidth}
        \includegraphics[width=\linewidth]{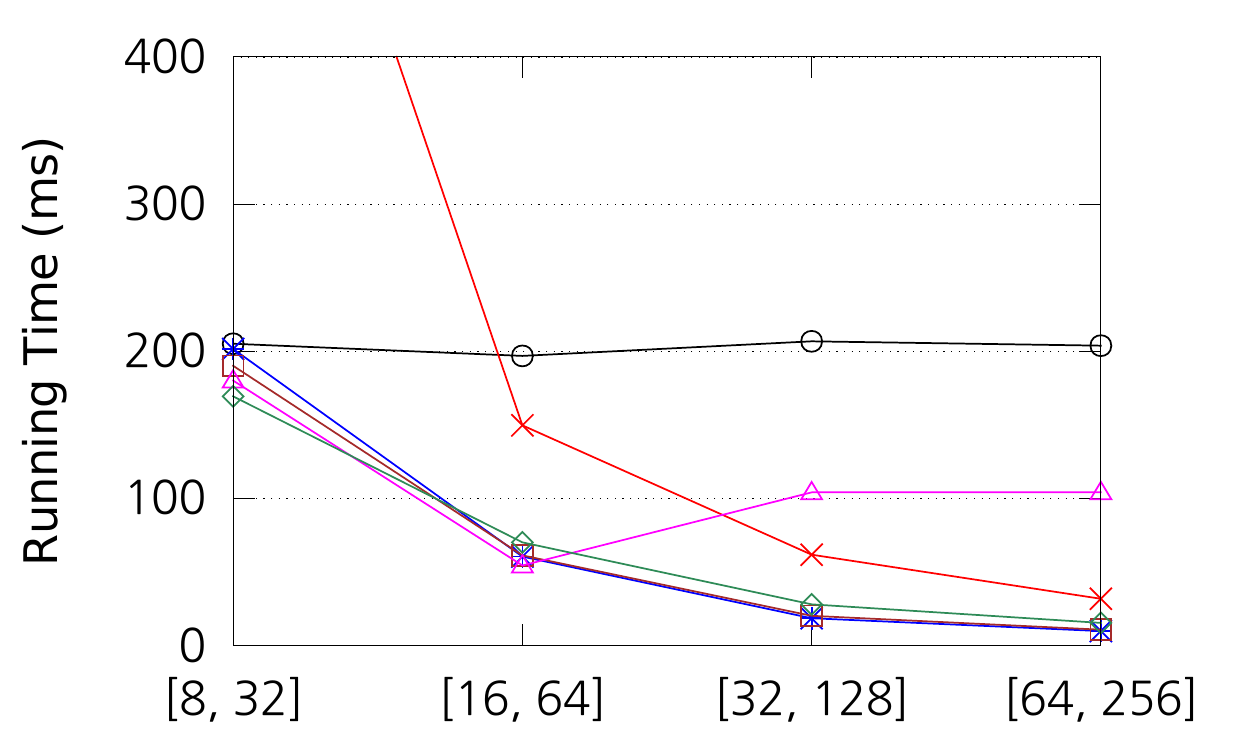}
        \vspace{-5mm}
        \caption{$k=50$}
        \label{fig:exp_diff_length_k50}
    \end{subfigure}
    
    \begin{subfigure}[b]{0.45\textwidth}
        \includegraphics[width=\linewidth]{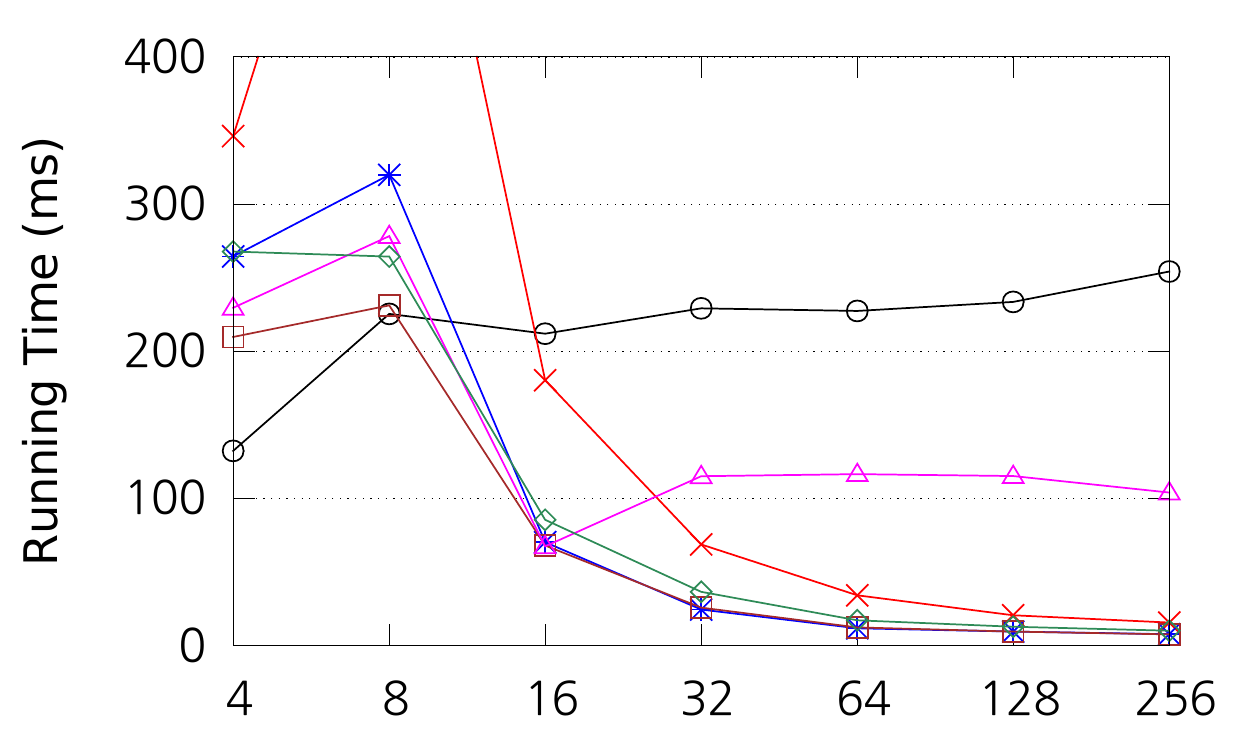}
        \vspace{-5mm}
        \caption{$k=100$}
        \label{fig:exp_equal_length_k100}
    \end{subfigure}
    \begin{subfigure}[b]{0.45\textwidth}
        \includegraphics[width=\linewidth]{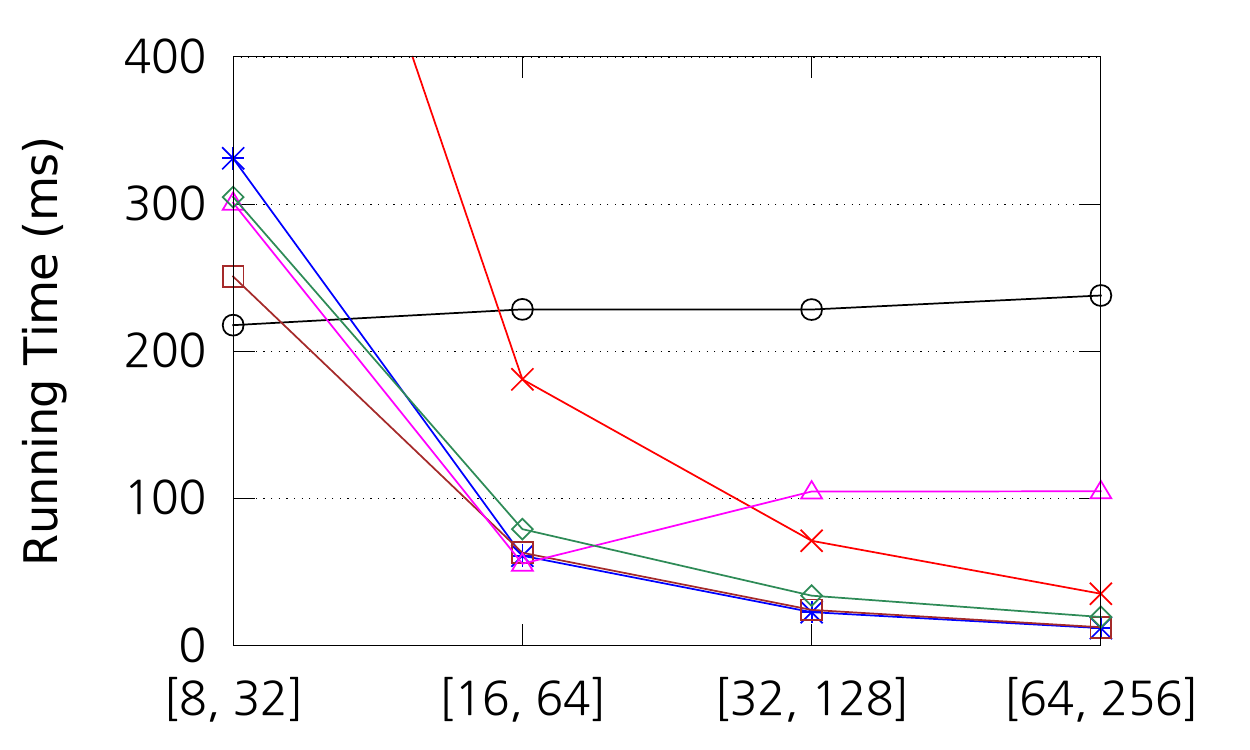}
        \vspace{-5mm}
        \caption{$k=100$}
        \label{fig:exp_diff_length_k100}
    \end{subfigure}
    
    \caption{Evaluation on the length of pattern. Left: patterns of equal length. Right: patterns of different lengths.}
    \label{fig:exp_pattern_length}
\end{figure}

        We first conduct experiments with sets of patterns of the same length. \red{Figures \ref{fig:exp_equal_length_k10}, \ref{fig:exp_equal_length_k50}, \ref{fig:exp_equal_length_k100}, and Table \ref{tab:equal}} show the results, where $k$ is the number of patterns and x-axis represents the length of the patterns, i.e., $m$. As the length of the patterns increases, WMB, WMBM, and ASB become the fastest algorithms due to a long shift length, low verification time, and light fingerprinting method. WMBM and WMB outperforms AC up to 33 times ($k=100$ and $m=256$). ASB outperforms AC up to 28 times ($k=10$ and $m=256$). RK outperforms AC up to 3 times ($k=50,100$ and $m=16$). When the length of the patterns is extremely short, however, AC is the clear winner ($m=4$). In this case, other algorithms na\"ively compare the greatest part of patterns for each position of the text. WMP works visibly worse when $m=8$ due to the extreme situation and overhead of the fingerprinting method. Since short patterns are more likely to have the same Cartesian trees, the proposed algorithms are sometimes faster when $m=4$ than when $m=8$ due to the grouping technique in Appendix \ref{sec:optimization}. Comparing WMB and WMBM, the min-index filtering method is more effective when there are many short patterns ($k=100$ and $m=4, 8$).
             
        \begin{table}[t]
            \centering
            \begin{tabular}{cccccccc}
                \toprule
                $k$ & $m$ & AC & WMP & WMB & WMBM & RK & ASB \\
                \midrule
                \multirow{6}{*}{10} & 4 & \textbf{129.46} & 303.093 & 176.249 & 166.04 & 165.351 & 147.889\\
                                    & 8 & 142.114 & 241.573 & 83.6087 & 92.1517 & \textbf{69.0761} & 88.5753\\
                                    & 16 & 138.79 & 93.5485 & \textbf{30.7575} & 33.4786 & 57.5673 & 39.4656\\
                                    & 32 & 160.921 & 42.6767 & \textbf{12.3674} & 13.3405 & 115.497 & 21.5187\\
                                    & 64 & 156.562 & 25.2625 & \textbf{7.59158} & 8.29616 & 115.381 & 11.0158\\
                                    & 128 & 145.905 & 15.0862 & \textbf{5.0663} & 5.97869 & 115.296 & 7.03843\\
                                    & 256 & 157.123 & 9.00974 & \textbf{4.69218} & 4.81503 & 102.995 & 5.43152\\
                \multirow{6}{*}{50} & 4 & \textbf{130.961} & 345.84 & 257.453 & 209.683 & 229.506 & 267.698\\
                                    & 8 & 203.431 & 651.249 & 193.496 & 173.894 & 181.898 & \textbf{150.484}\\
                                    & 16 & 197.931 & 145.531 & \textbf{58.6471} & 59.2581 & 63.8881 & 68.6459\\
                                    & 32 & 201.09 & 59.732 & \textbf{21.66} & 22.8856 & 115.723 & 30.24\\
                                    & 64 & 197.544 & 30.9735 & \textbf{9.86238} & 10.6876 & 115.721 & 14.7707\\
                                    & 128 & 203.944 & 18.0982 & \textbf{6.73188} & 6.9642 & 116.156 & 9.65942\\
                                    & 256 & 221.186 & 12.0733 & \textbf{6.57459} & 6.66625 & 103.055 & 8.05778\\
                \multirow{6}{*}{100} & 4 & \textbf{132.263} & 346.139 & 264.371 & 209.588 & 229.396 & 267.633\\
                                    & 8 & \textbf{225.327} & 681.149 & 319.767 & 231.097 & 278.165 & 264.218\\
                                    & 16 & 211.893 & 180.281 & 70.2239 & 67.93 & \textbf{67.3007} & 85.3792\\
                                    & 32 & 229.12 & 68.7025 & \textbf{24.4567} & 25.7314 & 115.032 & 36.4216\\
                                    & 64 & 227.275 & 34.1059 & \textbf{11.6273} & 12.3154 & 116.446 & 17.1575\\
                                    & 128 & 233.471 & 20.4809 & 9.49517 & \textbf{9.43364} & 115.08 & 12.6862\\
                                    & 256 & 254.042 & 15.563 & 7.66052 & \textbf{7.5831} & 103.943 & 9.98069\\
                \bottomrule
            \end{tabular}
            \caption{Evaluation on the patterns of equal length. Total time in ms.}
            \vspace{-6mm}
            \label{tab:equal}
        \end{table}
 
        \begin{table}[t!]
            \centering
            \begin{tabular}{cccccccc}
                \toprule
                $k$ & interval & AC & WMP & WMB & WMBM & RK & ASB \\
                \midrule
                \multirow{4}{*}{10} & [8, 32] & 152.628 & 240.46 & 97.506 & 103.019 & \textbf{65.6954} & 97.5208\\
                                    & [16, 64] & 153.663 & 95.9347 & \textbf{30.7831} & 33.076 & 50.4686 & 35.7311\\
                                    & [32, 128] & 150.329 & 44.4056 & \textbf{12.1087} & 13.629 & 103.051 & 19.3249\\
                                    & [64, 256] & 147.741 & 25.5997 & \textbf{7.22873} & 7.83777 & 102.949 & 10.1762\\
                \multirow{4}{*}{50} & [8, 32] & 205.042 & 724.675 & 201.416 & 190.008 & 180.04 & \textbf{169.276}\\
                                    & [16, 64] & 196.745 & 149.612 & 60.3754 & 61.1807 & \textbf{54.4075} & 70.1237\\
                                    & [32, 128] & 206.627 & 61.7051 & \textbf{18.5565} & 20.2259 & 104.028 & 27.9782\\
                                    & [64, 256] & 203.731 & 31.6943 & \textbf{9.79816} & 10.678 & 104.11 & 15.3719\\
                \multirow{4}{*}{100} & [8, 32] & \textbf{217.625} & 757.974 & 331.015 & 250.613 & 300.803 & 304.732\\
                                    & [16, 64] & 228.42 & 180.796 & 60.9719 & 63.0149 & \textbf{55.602} & 79.0707\\
                                    & [32, 128] & 228.194 & 71.0881 & \textbf{22.5928} & 24.1753 & 104.574 & 33.8765\\
                                    & [64, 256] & 237.803 & 35.1944 & \textbf{11.8472} & 12.4182 & 104.79 & 19.3238\\
                \bottomrule
            \end{tabular}
            \caption{Evaluation on the patterns of different lengths. Total time in ms.}
            \label{tab:diff}
        \end{table}
           
    \subsection{Evaluation on the Different Length Patterns}

        We compare algorithms with sets of patterns of different lengths. \red{Figures \ref{fig:exp_diff_length_k10}, \ref{fig:exp_diff_length_k50}, \ref{fig:exp_diff_length_k100}, and Table \ref{tab:diff}} show the results. The length is randomly selected in an interval, i.e., [8, 32], [16, 64], [32, 128], and [64, 256]. After a length is selected, a pattern is extracted from the text at a random position. 
        When there are many short patterns, i.e., $k=100$ and patterns of length 8--32, AC is the fastest due to the short minimum pattern length.
        
        When the length of the shortest pattern is sufficiently long, however, the proposed algorithms outperform AC. Specifically WMB outperforms AC up to 20 times ($k=10,50,100$ and patterns of length 64--256). ASB outperforms AC up to 14 times ($k=10$ and patterns of length 64--256). RK outperforms AC up to 4 times ($k=100$ and patterns of length 16--64).

    \subsection{Evaluation on the Real Dataset}
        
        \begin{figure}[t]
    \centering

    \begin{subfigure}[b]{0.9\textwidth}
        \includegraphics[width=\linewidth]{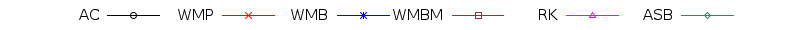}
        \label{fig:exp_pattern_length_label}
    \end{subfigure}

    \begin{subfigure}[b]{0.32\textwidth}
        \includegraphics[width=\linewidth]{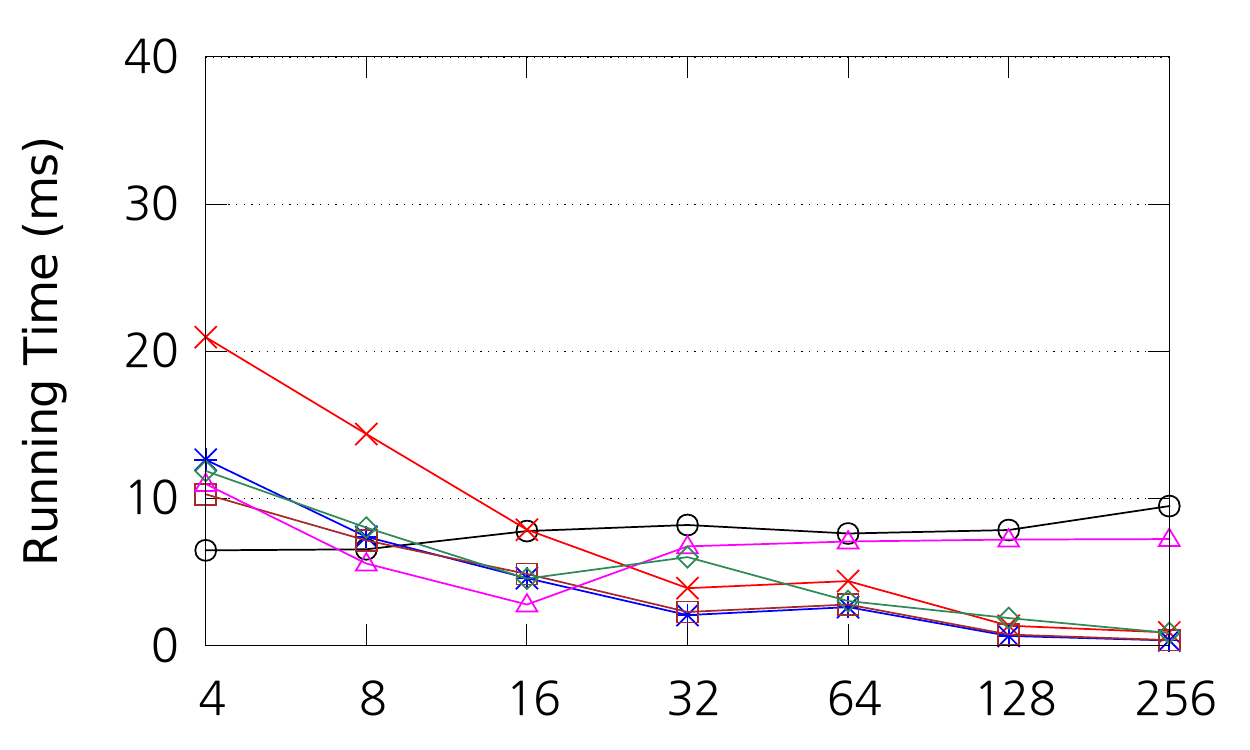}
        \caption{$k=10$}
        \label{fig:exp_real_k10}
    \end{subfigure}
    \begin{subfigure}[b]{0.32\textwidth}
        \includegraphics[width=\linewidth]{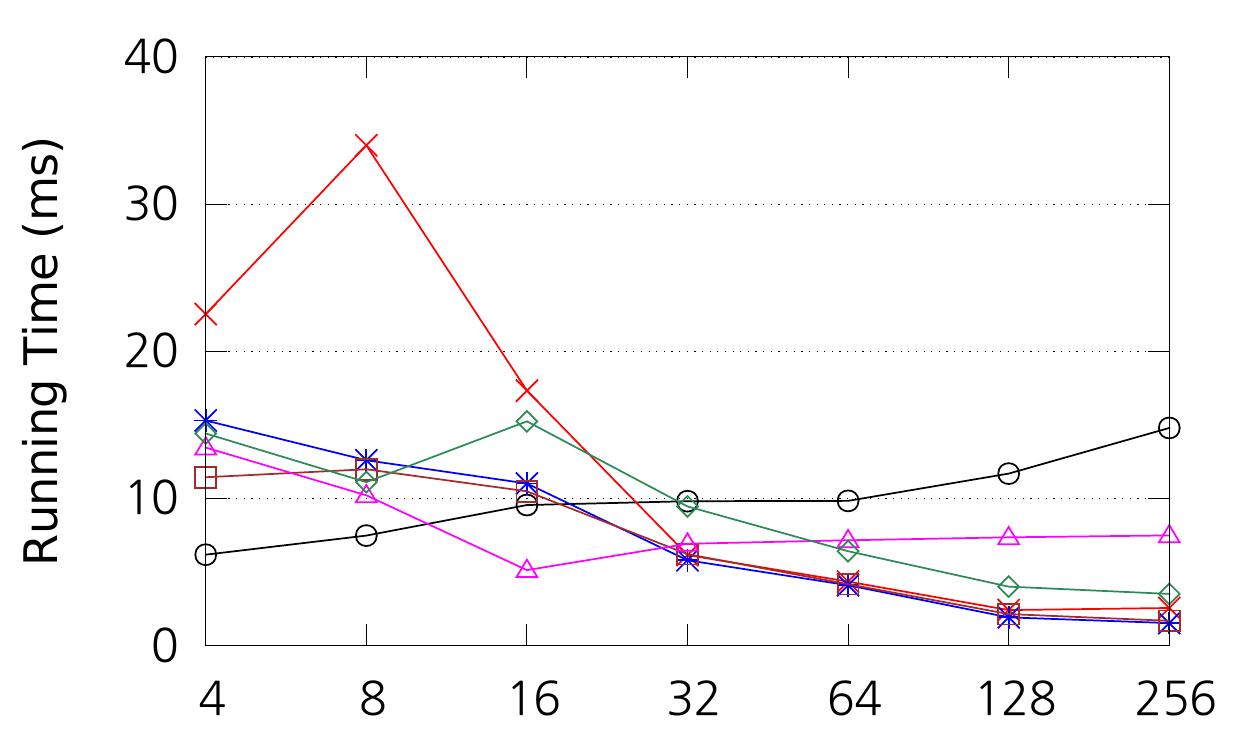}
        \caption{$k=50$}
        \label{fig:exp_int_k10}
    \end{subfigure}
    \begin{subfigure}[b]{0.32\textwidth}
        \includegraphics[width=\linewidth]{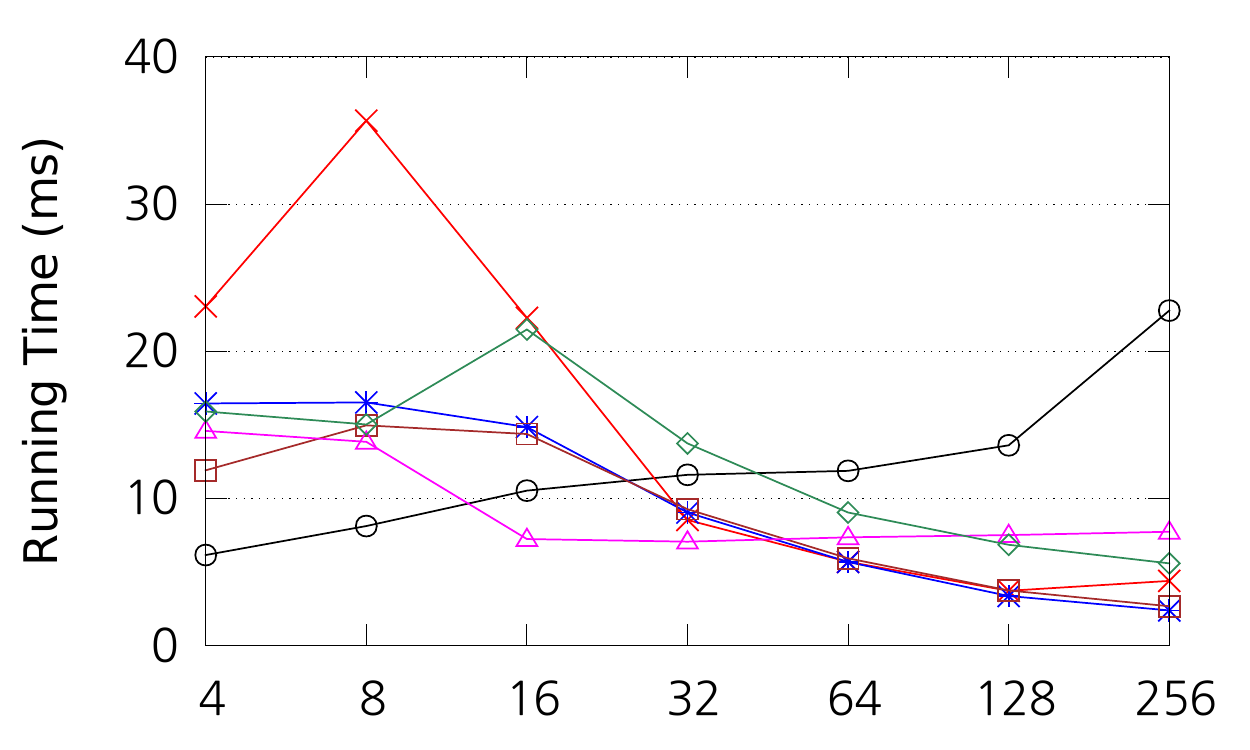}
        \caption{$k=100$}
        \label{fig:exp_char_k10}
    \end{subfigure}
    
    \caption{Evaluation on the Seoul temperatures dataset.}
    \label{fig:exp_seoul}
\end{figure}
    
        \begin{table}[b!]
            \vspace{3mm}
            \centering
            \begin{tabular}{cccccccc}
                \toprule
                $k$ & $m$ & AC & WMP & WMB & WMBM & RK & ASB \\
                \midrule
                \multirow{6}{*}{10} & 4 & \textbf{6.46631} & 20.9454 & 12.6187 & 10.2736 & 10.9732 & 11.8492\\
                                    & 8 & 6.53721 & 14.3666 & 7.37876 & 7.14195 & \textbf{5.57104} & 8.00697\\
                                    & 16 & 7.76917 & 7.8657 & 4.57646 & 4.85934 & \textbf{2.78754} & 4.5365\\
                                    & 32 & 8.18157 & 3.89075 & \textbf{2.06438} & 2.27235 & 6.73496 & 5.99976\\
                                    & 64 & 7.60696 & 4.37882 & \textbf{2.60346} & 2.7861 & 7.06377 & 3.01767\\
                                    & 128 & 7.84501 & 1.34436 & \textbf{0.643153} & 0.743147 & 7.19664 & 1.86238\\
                                    & 256 & 9.47242 & 0.88061 & \textbf{0.337183} & 0.36453 & 7.22575 & 0.850833\\
                \multirow{6}{*}{50} & 4 & \textbf{6.1634} & 22.5166 & 15.2899 & 11.4285 & 13.4452 & 14.4079\\
                                    & 8 & \textbf{7.47185} & 33.9852 & 12.581 & 11.9699 & 10.2026 & 11.0986\\
                                    & 16 & 9.53764 & 17.3211 & 11.0096 & 10.48 & \textbf{5.12495} & 15.234\\
                                    & 32 & 9.80261 & 6.14176 & \textbf{5.79404} & 6.21041 & 6.90745 & 9.44348\\
                                    & 64 & 9.82792 & 4.34029 & \textbf{4.09002} & 4.16979 & 7.15372 & 6.4055\\
                                    & 128 & 11.6782 & 2.40814 & \textbf{1.91395} & 2.1363 & 7.34409 & 3.99501\\
                                    & 256 & 14.7849 & 2.54673 & \textbf{1.5183} & 1.67897 & 7.47328 & 3.50649\\
                \multirow{6}{*}{100} & 4 & \textbf{6.15083} & 23.0344 & 16.4377 & 11.9024 & 14.58 & 15.904\\
                                     & 8 & \textbf{8.11009} & 35.6604 & 16.5101 & 14.9557 & 13.8331 & 15.015\\
                                     & 16 & 10.5246 & 22.2591 & 14.8361 & 14.3679 & \textbf{7.22885} & 21.4713\\
                                     & 32 & 11.5976 & 8.5304 & 9.03897 & 9.25709 & \textbf{7.05395} & 13.7257\\
                                     & 64 & 11.8653 & 5.6808 & \textbf{5.67174} & 5.92024 & 7.35357 & 9.04152\\
                                     & 128 & 13.6058 & 3.71476 & \textbf{3.36717} & 3.74349 & 7.50687 & 6.83653\\
                                     & 256 & 22.7509 & 4.3859 & \textbf{2.38758} & 2.66045 & 7.73048 & 5.58111\\
                \bottomrule
            \end{tabular}
            \caption{Evaluation on the Seoul temperatures dataset. Total time in ms.}
            \label{tab:seoul}
        \end{table}    
        
        We conduct experiment on a real dataset, which is a time series of Seoul temperatures. The Seoul temperatures dataset consists of 658,795 integers referring to the hourly temperatures in Seoul (multiplied by ten) in the years 1907--2019 \cite{song2019fast}. In general, temperatures rise during the day and fall at night. Therefore, the Seoul temperatures dataset has more matches than random datasets when patterns are extracted from the text. \red{Figure \ref{fig:exp_seoul} and Table \ref{tab:seoul} show} the results on the Seoul temperatures dataset with sets of patterns of the same length. As the pattern length grows, the proposed algorithms run much faster than AC. For short patterns ($m=4,8$), AC is the fastest algorithm, and AC is up to twice times faster than WMBM ($m=4$ and $k=100$) and 1.7 times faster than RK ($m=8$ and $k=100$). For moderate-length patterns ($m=16,32$), RK is up to 2.8 times faster than AC ($m=16$ and $k=10$), and WMB is up to 4 times faster than AC ($m=32$ and $k=10$). For relatively long patterns ($m=64,128,256$), all the proposed algorithms outperform AC. Specifically, WMB, WMBM, ASB, and WMP outperform AC up to 28, 26, 11, and 10 times, respectively ($m=256$ and $k=10$), and RK outperforms AC up to 2.9 times ($m=256$ and $k=100$).
        

\FloatBarrier 
\bibliographystyle{splncs04}
\bibliography{bibliography}

\begin{thebibliography}{10}
\providecommand{\url}[1]{\texttt{#1}}
\providecommand{\urlprefix}{URL }
\providecommand{\doi}[1]{https://doi.org/#1}

\bibitem{aho1975efficient}
Aho, A.V., Corasick, M.J.: Efficient string matching: an aid to bibliographic
  search. Communications of the ACM  \textbf{18}(6),  333--340 (1975)

\bibitem{berkman1993optimal}
Berkman, O., Schieber, B., Vishkin, U.: Optimal doubly logarithmic parallel
  algorithms based on finding all nearest smaller values. Journal of Algorithms
   \textbf{14}(3),  344--370 (1993)

\bibitem{boyer1977fast}
Boyer, R.S., Moore, J.S.: A fast string searching algorithm. Communications of
  the ACM  \textbf{20}(10),  762--772 (1977)

\bibitem{charras1998very}
Charras, C., Lecroq, T., Pehoushek, J.D.: A very fast string matching algorithm
  for small alphabets and long patterns. In: Annual Symposium on Combinatorial
  Pattern Matching. pp. 55--64. Springer (1998)

\bibitem{chhabra2014order}
Chhabra, T., Tarhio, J.: Order-preserving matching with filtration. In:
  International Symposium on Experimental Algorithms. pp. 307--314. Springer
  (2014)

\bibitem{commentz1979string}
Commentz-Walter, B.: A string matching algorithm fast on the average. In:
  International Colloquium on Automata, Languages, and Programming. pp.
  118--132. Springer (1979)

\bibitem{cormen2001introduction}
Cormen, T.H., Leiserson, C.E., Rivest, R.L., Stein, C.: Introduction to
  algorithms second edition. The Knuth-Morris-Pratt Algorithm  (2001)

\bibitem{crochemore1999fast}
Crochemore, M., Czumaj, A., Gasieniec, L., Lecroq, T., Plandowski, W., Rytter,
  W.: Fast practical multi-pattern matching. Information Processing Letters
  \textbf{71}(3-4),  107--113 (1999)

\bibitem{ganguly2016space}
Ganguly, A., Hon, W.K., Sadakane, K., Shah, R., Thankachan, S.V., Yang, Y.:
  Space-efficient dictionaries for parameterized and order-preserving pattern
  matching. In: 27th Annual Symposium on Combinatorial Pattern Matching (CPM).
  pp. 2:1--2:12. LIPIcs (2016)

\bibitem{han2015fast}
Han, M., Kang, M., Cho, S., Gu, G., Sim, J.S., Park, K.: Fast multiple
  order-preserving matching algorithms. In: International Workshop on
  Combinatorial Algorithms. pp. 248--259. Springer (2015)

\bibitem{hua2009variable}
Hua, N., Song, H., Lakshman, T.: Variable-stride multi-pattern matching for
  scalable deep packet inspection. In: IEEE INFOCOM 2009. pp. 415--423. IEEE
  (2009)

\bibitem{karp1987efficient}
Karp, R.M., Rabin, M.O.: Efficient randomized pattern-matching algorithms. IBM
  journal of research and development  \textbf{31}(2),  249--260 (1987)

\bibitem{kim2014order}
Kim, J., Eades, P., Fleischer, R., Hong, S.H., Iliopoulos, C.S., Park, K.,
  Puglisi, S.J., Tokuyama, T.: Order-preserving matching. Theoretical Computer
  Science  \textbf{525},  68--79 (2014)

\bibitem{knuth2014art}
Knuth, D.E.: The Art of Computer Programming, volume 2: Seminumerical
  algorithms. Addison-Wesley Professional (2014)

\bibitem{kubica2013linear}
Kubica, M., Kulczy{\'n}ski, T., Radoszewski, J., Rytter, W., Wale{\'n}, T.: A
  linear time algorithm for consecutive permutation pattern matching.
  Information Processing Letters  \textbf{113}(12),  430--433 (2013)

\bibitem{liao2013intrusion}
Liao, H.J., Lin, C.H.R., Lin, Y.C., Tung, K.Y.: Intrusion detection system: A
  comprehensive review. Journal of Network and Computer Applications
  \textbf{36}(1),  16--24 (2013)

\bibitem{liu2007automatic}
Liu, J.N., Kwong, R.W.: Automatic extraction and identification of chart
  patterns towards financial forecast. Applied Soft Computing  \textbf{7}(4),
  1197--1208 (2007)

\bibitem{park2019cartesian}
Park, S., Amir, A., Landau, G.M., Park, K.: Cartesian tree matching and
  indexing. In: 30th Annual Symposium on Combinatorial Pattern Matching (CPM).
  pp. 16:1--16:14. LIPIcs (2019)

\bibitem{song2019fast}
Song, S., Ryu, C., Faro, S., Lecroq, T., Park, K.: Fast cartesian tree matching
  algorithms. Accepted to SPIRE (2019), https://arxiv.org/abs/1908.04937

\bibitem{song2008memory}
Song, T., Zhang, W., Wang, D., Xue, Y.: A memory efficient multiple pattern
  matching architecture for network security. In: IEEE INFOCOM 2008-The 27th
  Conference on Computer Communications. pp. 166--170. IEEE (2008)

\bibitem{vuillemin1980unifying}
Vuillemin, J.: A unifying look at data structures. Communications of the ACM
  \textbf{23}(4),  229--239 (1980)

\bibitem{wu1994fast}
Wu, S., Manber, U.: A fast algorithm for multi-pattern searching. Technical
  report. TR-94-17, Department of Computer Science, University of Arizona
  (1994)

\end{thebibliography}

\appendix
\section{APPENDIX}
\vspace{-5mm}
\begin{algorithm}[H]
        \caption{Algorithm based on Rabin-Karp}
        \label{alg:Rabin-Karp}
        \begin{algorithmic}[1]
            \State $\textbf{input: }\text{text } T[1..n]\text{ and patterns } P_1[1..m_1], P_2[1..m_2], ..., P_k[1..m_k]$
            \State $\textbf{output: }\text{every position in } T \text{ that matches at least one of the patterns}$
            
            \Procedure{Preprocessing}{}
                \State $m \gets \text{min}(m_1,m_2,...,m_k)$
                \For{$i \gets 1$ to $k$}
                    \State {$\fp$ $\gets$ $f(P_i[1..m])$}
                    \State $\HASH[\fp].add(i)$
                \EndFor
            \EndProcedure
                
            \Procedure{Search}{}
                \State $\idx \gets m$
                \While{$\idx \leq n$}
                    \State $\fp \gets f(T[\idx-m+1..index])$
                    \For{$i \in \HASH[\fp]$}
                        \If{$P_i$ matches $T[\idx-m+1..\idx-m+m_i]$}
                            \State output $\idx-m+m_i$
                        \EndIf
                    \EndFor
                    
                    \State $\idx \gets \idx + 1$
                \EndWhile
                
            \EndProcedure
        \end{algorithmic}
    \end{algorithm}
    
    \vspace{-10mm}
    
    \begin{algorithm}[H]
        \caption{Algorithm based on Alpha Skip Search}
        \label{alg:AlphSkipSearch}
        \begin{algorithmic}[1]
            \State $\textbf{input: }\text{text } T[1..n]\text{ and patterns } P_1[1..m_1], P_2[1..m_2], ..., P_k[1..m_k]$
            \State $\textbf{output: }\text{every position in } T \text{ that matches at least one of the patterns}$
            
            \Procedure{Preprocessing}{}
                \State $m \gets \text{min}(m_1,m_2,...,m_k)$
                \State $b \gets {\log_2 (km)}$
                \For{$i \gets 1$ to $k$}
                    \For{$j \gets b$ to $m$}
                        \State {$\fp$ $\gets$ $f(P_i[j-b+1..j])$}
                        \State $\POS[\fp].add(i,j)$
                    \EndFor
                \EndFor
            \EndProcedure
                
            \Procedure{Search}{}
                \State $\idx \gets m$
                \While{$\idx \leq n$}
                    \State $fp \gets f(T[\idx-b+1..\idx])$
                    \For{$(i,j) \in \POS[\fp]$}
                        \If{$P_i$ matches $T[\idx-j+1..\idx-j+m_i]$}
                            \State output $\idx-j+m_i$
                        \EndIf
                    \EndFor
                    \State $\idx \gets \idx + m-b+1$
                \EndWhile
            \EndProcedure
        \end{algorithmic}
    \end{algorithm}

\subsection{Proof of Lemma \ref{lem:matchingprobability}}
    \begin{proof}
        When the $i$-th numbers are the roots of both $CT(S_1)$ and $CT(S_2)$, the probability that $CT(S_1)=CT(S_2)$ is ${p(i-1)p(n-i)}$. Since there are $n$ distinct numbers, the probability that both $CT(S_1)$ and $CT(S_2)$ have the $i$-th numbers as their roots is ${1\over{n^2}}$. Summing the probabilities ${p(i-1)p(n-i) \over{n^2}}$ for $1 \leq i \leq n$ gives the probability $p(n)$. \qed
    \end{proof}

\subsection{Proof of Theorem \ref{thm:matchingprobability}}
    \begin{proof}
        We prove the theorem by induction on $n$.
        
        If $n \leq 3$, $p(1) = 1 \leq {1\over2^{0}}$, $p(2) = {1\over2} \leq {1\over2^{1}}$, 
        $p(3) = {2\over9} \leq {1\over2^{2}}$. Therefore, the theorem holds when $n\leq3$.
        
        Let's assume that the theorem holds when $n \leq k$, for $k\geq3$, and show that it holds when $n=k+1$.
        \begin{equation}
            \begin{split}
            p(k+1) &={p(0)p(k)+p(1)p(k-1)+\dots+p(k)p(0) \over{(k+1)^2}} \\
            &\leq ({1\over2^{-1}} {1\over2^{k-1}} + {1\over2^{0}} {1\over2^{k-2}} + \dots + {1\over2^{k-1}} {1\over2^{-1}}) {1 \over {(k+1)^2}}\\
            &= {{k+1}\over {2^{k-2}}} {1 \over {(k+1)^2}}\\
            &\leq {1\over {2^{k-2}}} {1 \over {4}}\\
            &= {1\over {2^{k}}}
            \end{split}
        \end{equation}
        
        Therefore, we have proved that $p(n)\leq {1\over{2^{n-1}}}$. \qed
    \end{proof}

    \subsection{Optimization Techniques}
        \label{sec:optimization}
    \subsubsection{Optimizing Na\"ive Verification}
    \label{subsec:verification}
        An efficient verification method is essential for the proposed three algorithms because they all adopt the filtering and verification approach. We employ the verification method introduced by Song et al.~\cite{song2019fast}. They first introduce the notion of the \textit{global-parent representation} $\mathcal{GP}_S[1..m]$ of a string $S[1..m]$, where $\mathcal{GP}_S[i]$ stores the index of the parent of $S[i]$ in $CT(S[1..m])$. For example, the global-parent representation of string $S=(11,14,13,15,12)$ is $\mathcal{GP}_S=(1,3,5,3,1)$. Note that the parent of the root is the root itself. Two strings $S_1[1..m]$ and $S_2[1..m]$ have the same Cartesian trees if and only if $S_1[\mathcal{GP}_{S_2}[i]] < S_1[i]$, or $S_1[\mathcal{GP}_{S_2}[i]] = S_1[i]$ with $\mathcal{GP}_{S_2}[i] \leq i$, for all $1 \leq i \leq m$ \cite{song2019fast}. Note that we do not need any representation of $S_1$. After the global-parent representation of $S_2$ is computed, we can verify whether $CT(S_1)=CT(S_2)$ in linear time by checking the conditions. In our algorithms, the global-parent representation of the patterns are computed and stored in advance, and verification is done by the above method without computing any representation about the text.
        
    \subsubsection{Reusing Fingerprint of Binary Encoding}
    \label{subsec:reusefingerprint}
        In Algorithm \ref{alg:Rabin-Karp}, successive fingerprints can be computed in constant time by Equation (\ref{eq:successivefinger}) when using the binary encoding. Likewise, we can reuse a previous fingerprint to create the current fingerprint in Algorithms \ref{alg:Wu-Manber} and \ref{alg:AlphSkipSearch} as well. This can be done by applying Equation (\ref{eq:successivefinger}) $b-l$ times when two blocks of size $b$ overlap by $l$. Our experimental study showed that reusing fingerprints when $l=b-1$ is the most efficient. Thus, we reuse fingerprints only when the text shifts by one position. It is worth mentioning that we do not reuse fingerprints of the parent-distance encoding because multiple characters in the parent-distance representation can be changed by just one shift, countervailing the effect of reusing.
        
    \subsubsection{Additional Filtering via Min-index}
    \label{subsec:minfiltering}
        In the filtering stage of an algorithm, we may further filter out candidate patterns by additional filtering methods. We introduce a simple filtering method based on the index of the minimum value (\emph{min-index}). Since two strings have the same Cartesian trees only if the indices of the minimum values (roots) of the two strings are the same, we may first compare the min-index before we verify each candidate pattern retrieved by a fingerprint. To this end, for each input pattern $P$, we store the min-index among $P[m-b+1..m]$ where $m$ is the length of the shortest pattern in the preprocessing step. In the search step, the fingerprint and the min-index of a block in the text are computed at the same time. Among the patterns retrieved by the fingerprint, only patterns $P_i$ are verified such that the min-index of the last block in $P_i$'s length-$m$ prefix is the same as that of the block in the text.
        The information of the root is not represented by the binary representation, but it is represented by the parent-distance representation. Therefore, this additional filtering method is effective only when we use the binary encoding.
        
    \subsubsection{Grouping Patterns Having the Same Cartesian Trees}
        Since the input patterns are strings, some of them may have the same Cartesian trees. The Aho-Corasick algorithm \cite{park2019cartesian} assembles those patterns in a state of its automaton, while the presented algorithms in this paper do not perform it explicitly. In our implementation, we group those patterns having the same Cartesian trees, so as to avoid the redundant computation. This process is particularly beneficial for short input patterns. 

\end{document}